\documentclass[11pt]{article} 
\usepackage{latexsym}
\usepackage{a4}
\usepackage[dvips]{graphicx}
\usepackage{bbm}      
\newcommand{\tr}{\mbox{tr}}

\newcommand{\unit}{\mathbbm{1}}

\newcommand{\Vek}[1]{\mbox{\boldmath$#1$\unboldmath}}
\newcommand{\vek}[1]{\mathbf{#1}}
\newcommand{\MM}{\mathsf{M}}
\newcommand{\Real}{\mathbbm{R}}

\newcommand{\Integer}{\mathbbm{Z}}

\newcommand{\GG}{\mathsf{G}}
\newcommand{\NN}{N}
\newcommand{\WW}{W}
\newcommand{\XX}{\mathsf{X}}
\newcommand{\DD}{\mathsf{D}}
\newcommand{\Me}{\overline{\mathsf{M}}_\epsilon}
\newcommand{\len}{n}

\begin{document} 
\begin{titlepage}
\mbox{ UNITU-THEP-17/1998}
  \vspace{1.5cm}


\renewcommand{\thefootnote}{\fnsymbol{footnote}} 

\begin{center}
  \Large\bfseries
   Magnetic Monopoles and Topology of  Yang-Mills Theory in Polyakov 
   Gauge
\end{center}
\vspace{1cm}


\renewcommand{\thefootnote}{\fnsymbol{footnote}} 
\centerline{M.~Quandt\footnote[1]{Supported by "Graduiertenkolleg: Hadronen und 
           Kerne"}, H.~Reinhardt, A.~Sch\"afke${}^\ast$}
\centerline{ Institut f\"ur Theoretische Physik, Universit\"at T\"ubingen }
\centerline{D--72076 T\"ubingen, Germany.}  \vspace{1.5cm}
\renewcommand{\thefootnote}{arabic{footnote}} \vspace{1.5cm}


\begin{abstract}
  We express the Pontryagin index in Polyakov gauge completely in 
  terms of magnetically charged gauge fixing defects, namely magnetic 
  monopoles, lines, and domain walls. Open lines and domain walls 
  are topologically equivalent to monopoles, which are the genuine defects.
  The emergence of non-genuine magnetically charged closed domain walls can be 
  avoided by choosing the temporal gauge field smoothly. The Pontryagin index 
  is then exclusively determined by the magnetic monopoles.
\end{abstract}

\vspace{2cm}

\end{titlepage}


\section{Introduction} 
\label{sec:1}
Recent lattice calculations \cite{bali} give evidence that confinement is 
realized as a dual Meissner effect, at least in the so-called Abelian gauges 
\cite{thooft}. In these gauges magnetic monopoles arise as obstructions to 
fixing the coset $G/H$ of the gauge group $G$, where $H$ is the Cartan subgroup 
which is left invariant. Lattice calculations indicate that these monopoles 
are in fact condensed \cite{giacomo}, a necessary condition for the QCD vacuum 
forming a dual superconductor.

Magnetic monopoles  are long ranged fields and should hence contribute to the 
topological properties of gauge fields. Furthermore topologically non-trivial
field configurations can explain spontaneous breaking of chiral 
symmetry \cite{cash}.
It is therefore interesting to clarify to which extent magnetic monopoles 
contribute to the topology of gauge fields.

Previously it was shown by one of us that in the Polyakov gauge, 
\begin{equation}
\Omega(\vek{x}) = \mathsf{P}\exp\left(-\int d x_0\,A_0\right) \stackrel{!}{=}
\mbox{diag}
\label{G0}\,,
\end{equation}
which is a particular Abelian gauge, magnetic monopoles completely account 
for the non-trivial topology of Yang-Mills fields.
To be specific, consider a pure Yang-Mills theory with colour group $SU(2)$. 
In Polyakov gauge the magnetic monopoles arise at those points in three-space 
where the Polyakov loop $\Omega (\vek{x})$ becomes an irregular element of the 
gauge group, $\Omega (\vek{x}_{i})=(-\unit)^{n_i}=\pm \unit$, i.e.~a centre
element for $SU(2)$. Here $\len_{i}$ is an integer. Furthermore, in order for 
the Pontryagin index to be well defined and the action finite, the gauge fields 
have to become asymptotically a pure gauge, which in turn implies that the 
Polyakov loop approaches an angle independent value at spatial infinity,
\begin{equation}
\lim\limits_{r\to\infty}\Omega(r,\hat{\vek{x}})=(-\unit)^{n_0}
\label{A1}\,.
\end{equation}
The following expression was derived for the Pontryagin index \cite{rei}
\begin{equation}
\nu=-\sum_i\ell_i m_i
\label{A2}\,,
\end{equation}
where $m_i$ denotes the magnetic charge of the monopole and $\ell_i=n_i-n_0$ is
an integer which can be interpreted as the invariant length traced out by the
Dirac string in group space. 

The boundary condition (\ref{A1}) allows us to compactify our spatial manifold
\begin{equation}
\Real^3\to\dot{\Real}^3 = \Real^3 \cup \{\infty\}\simeq S^3
\label{A3}\,.
\end{equation}
In this way, the surface at spatial infinity becomes a point of the compactified
manifold, $\dot{\Real}^3\simeq S^3$, which hosts a magnetic monopole due to the
b.c.~(\ref{A1}). Furthermore on a compact manifold the net magnetic charge of 
all monopoles has to vanish, $\sum_{i} m_{i}=0$. Given this setting, 
the expression for the Pontryagin index found in ref.~\cite{rei} simplifies to 
\begin{equation}
\nu = - \sum\limits_i m_i\,\len_i
\label{G1}
\end{equation}
where the summation is now over all magnetic monopoles including the one at 
the infinitely distant point.

Subsequently the same problem has been treated in a somewhat different fashion 
in references \cite{wipf,lenz} resulting in the following expression for the 
Pontryagin index,
\begin{equation}
\nu = - \sum\limits_{i\atop{\Omega = - \unit}} m_i\,,
\label{G3}
\end{equation} 
where the summation is performed over the charges $m_i$ of magnetic monopoles 
corresponding to the irregular element $\Omega=-\unit$ while in  
eq.~(\ref{G1}) the summation is over all monopoles. In addition the invariant 
length of the Dirac string enters only in eq.~(\ref{G1}). Formally 
eq.~(\ref{G3}) results from eq.~(\ref{G1}) by restricting the integers $n_k$ to 
$n_k=0,1$. 

In the present paper 
we will summarize the result of a thorough investigation of the topological 
charge in the presence of gauge fixing defects. In particular we will show 
that eq.~(\ref{G1}) is more general than eq.~(\ref{G3}).
While the latter formula gives the correct winding number only in the absence 
of domain walls, eq.~(\ref{G1}) includes already the effect of non-genuine 
domain walls, which arise when $\ln\Omega$ is restricted to first Weyl 
alcove.\footnote{A Weyl alcove is a fundamental domain in the Cartan algebra
with respect to the extended Weyl group, i.e.~any of the discrete symmetries
(displacements or Weyl symmetries) leads out of the alcove. For
$G=SU(2)$ as an example, the Cartan group is $\{e^{i\chi\sigma_3}\}$, the
displacements are $\chi\to\chi +2\pi m$ and the Weyl symmetry is $\chi\to-\chi$
whence the Weyl alcove is found to be $\chi\in [0,\pi]$.}


\section{Abelian Gauge Fixing}
\label{sec:2}

The starting point of the (canonical) quantization of Yang-Mills theory is the 
Weyl-gauge \cite{jackiw}
\begin{equation}
A_0 = 0 \,.
\label{G4}
\end{equation}
It is generally assumed that in this gauge the dynamical fields, i.e.~the 
spatial field components $A_{i=1,2,3}(x)$ are smooth functions of space-time. 
The quantity of interest is the gauge invariant partition function 
for which it is straightforward to derive the following functional integral 
representation \cite{johns,hugo}: 
\begin{equation}
Z = \int\limits_G \mathcal{D}\mu[\Omega]\cdot e^{-i n[\Omega]\theta} 
\,\int\limits_{{\rm b.c.\,}(\Omega)}\mathcal{D}\vek{A}\,
\exp\left(- S_{\rm YM}[A_0 = 0,\vek{A}]\right)
\label{G5} \,.
\end{equation}
Gauge invariance requires here the spatial gauge fields to satisfy the twisted 
boundary condition 
\begin{equation}
\vek{A}(t=0,\vek{x}) = \left(\vek{A}(t = \beta,\vek{x})\right)^\Omega
\label{G6},
\end{equation} 
where $A^{\Omega}=\Omega\,A\,\Omega^{\dagger}+\Omega\,\partial\,\Omega^\dagger$
is the gauge transformed field, and further requires to integrate over all gauge functions 
$\Omega (\vek{x})$ with the invariant (Haar) measure $\mu$. Like the dynamical fields 
$A_{i=1,2,3}(x)$, the gauge rotation $\Omega (\vek{x})$ can be assumed to be 
smooth.

Topologically the gauge fields are classified by the Pontryagin index 
\begin{equation}
\nu[A] = - \frac{1}{8 \pi^2} \int\limits_{\mathcal{M}} \tr\,
(F\wedge F) = \frac{1}{32\pi^2}\int\limits_{\mathcal{M}} d^4x\,
F_{\mu\nu}^a {}^\ast F_{\mu\nu}^a
\label{G7}\,,
\end{equation}
where $\mathcal{M}$ is the space-time manifold which we choose to be 
$\mathcal{M}=[0,\beta]\times \mathsf{M}$ with the spatial 
three-manifold $\mathsf{M}$. We shall specify $\mathsf{M}$ as the one-point 
compactification $\mathsf{M}=\Real^3 \cup \{\infty\}=\dot{\Real}^3\simeq S^3$.

In the Weyl gauge and with the twisted boundary condition (\ref{G6}) the 
Pontryagin index is given by the winding number 
\begin{equation}
n[\Omega] = -\frac{1}{24 \pi^2} \int\limits_\mathsf{M} \tr\,
\left(L\wedge L \wedge L\right) \,\, ; \qquad L = \Omega\cdot d\Omega^{-1}
\label{G8}
\end{equation}
of the gauge function $\Omega(\vek{x})$, i.e. 
\begin{equation}
\nu[A_0=0,\vek{A}] = n[\Omega]
\label{G9}\,.
\end{equation}
Note that this number enters with the vacuum angle $\theta$ in the partition
function (\ref{G5}).

For many purposes the twisted boundary conditions are inconvenient and it 
is useful to convert them to periodic ones, 
\begin{equation}
\vek{A}(t=\beta,\vek{x}) = \vek{A}(t=0,\vek{x})
\label{G10} \,,
\end{equation}
by performing the following time-dependent gauge transformation 
\cite{johns,hugo}
\begin{equation}
U(t,\vek{x}) = \Omega(\vek{x})^{\frac{t}{\beta}-1}
\label{G11}
\end{equation} 
which introduces a time-independent temporal gauge field component
\begin{equation}
A'_0 = (A_0 = 0)^U = U \partial_0 U^{-1} = - \beta^{-1}\,\ln\Omega(\vek{x})
\label{G12}\,.
\end{equation}
From the gauge defined by the previous equation, which is equivalent to
\begin{equation}
\partial_0 A_0 = 0
\label{G14}\,,
\end{equation} 
one arrives at the \emph{Polyakov gauge} by a time-independent gauge 
transformation
$V(\vek{x})$ diagonalizing $\Omega(\vek{x})$ and hence $A_0$,
\begin{equation}
\Omega(\vek{x}) = V^{-1}(\vek{x})\cdot\omega(\vek{x})\cdot V(\vek{x})
\stackrel{V}{\longrightarrow} \omega(\vek{x})
\label{G15}
\end{equation}
where $\omega$ and $V$ live in the Cartan subgroup $H\subset G$ and in the 
coset $G/H$, respectively. The coset element $V\in G/H$ which diagonalizes 
$\Omega \in G$ is obviously defined only up to an element of the
normalizer $\NN(H)$ of $H$ in $G$,
\begin{equation}
V \to g\cdot V\,\,,\qquad g\in \NN(H)
\label{G16}\,.
\end{equation}
The normalizer $\NN(H)$ is related to the Cartan subgroup $H$ by 
$\NN(H)=\WW\times H$ where $\WW$ denotes the Weyl group. For the gauge group 
$G=SU(N)$ the Weyl group $\WW$ is isomorphic to the permutation group $S_N$. 
In fact, the Weyl transformations $w\in\WW$ permute the diagonal elements of 
$\omega$ and are not part of the Cartan subgroup.

Topological obstructions to implementing the Polyakov 
gauge occur, and these are of three different types: 
\begin{enumerate}
\item The gauge function $\Omega(\vek{x})$ may take values corresponding to
      irregular elements of the 
      gauge group in which two eigenvalues coincide and the diagonalization, 
      i.e.~the coset element $V\in G/H$, is not well defined.\footnote{For the 
      gauge group $G=SU(2)$ the irregular elements coincide with the centre 
      elements $\pm\unit$.} In this case we have local gauge fixing defects, 
      which manifest themselves as magnetic charges in the induced gauge field
      $\mathcal{A}_i=V\partial_i V^\dagger$.
\item In the diagonalization $\Omega =V^{-1}\omega V$ the elements $\omega$ and 
      $V$ may not be globally defined and smooth on $\MM$ even if 
      $\Omega(\vek{x})$ is smooth and everywhere regular. The point is that 
      the compactification imposes certain boundary conditions on 
      $\Omega(\vek{x})$, and $\omega$ or $V$ may fail to obey these conditions.
\item Functions of matrices like $\Omega(\vek{x})$  are generally defined by 
      the spectral theorem, i.e.~even if diagonalization problems on $\MM$ are 
      absent, $f(\Omega)$ can only be smooth if the function $f$ is holomorphic 
      on the spectrum of $\Omega(\vek{x})$. For the fractional power in 
      eq.~(\ref{G11}) or the logarithm in eq.~(\ref{G12}) this may be 
      impossible due to the branch cut of the logarithm in the complex plane.
\end{enumerate}
The second type of obstructions has been discussed in ref.~\cite{blau}
and we will not consider it here. Furthermore we will see below that the third 
type of obstruction is automatically resolved by a proper treatment of the
defects arising in item one, and hence we will concentrate in the following on 
the investigation of the local gauge fixing defects. We will in particular show 
that in the Polyakov gauge the Pontryagin index arises entirely from these 
gauge defects.

Note that the gauge potentials in both the gauge (\ref{G14}) and the 
Polyakov gauge (\ref{G15}) fulfill the periodic boundary conditions (\ref{G10}) 
and thus live on the closed compact manifold 
$\mathcal{M} = S^1\times\mathsf{M}$. It is then easy to see that the 
Pontryagin index arises exclusively from the defects. 
As is well known, the integrand in the Pontryagin index is a total derivative 
for non-singular gauge fields, 
\begin{equation}
\tr\,(F\wedge F) = d\,\mathsf{K}[A]
\label{G18}
\end{equation}
where
\begin{equation}
\mathsf{K} = \tr\,\left(F\wedge A - \frac{1}{3} A \wedge A \wedge A\right)
\label{G19}
\end{equation} 
is the topological current. From $\partial\mathcal{M} = 0$ and Stokes' theorem 
we find 
\begin{equation}
\nu = -\frac{1}{8\pi^2}\int\limits_{\mathcal{M}} d\,\mathsf{K} 
= -\frac{1}{8\pi^2}\int\limits_{\partial \mathcal{M}}\mathsf{K} = 0
\label{G20}\,.
\end{equation}
The crucial observation is here that the gauge fixing defects for which 
the coset element $V\in G/H$ is ill-defined give rise to singular 
connections $A^V$ in Polyakov gauge (cf.~eq.~(\ref{G15})), so that at the 
gauge defects equation (\ref{G18}) does not apply.

A similar conclusion may be drawn from the results of ref.~\cite{rei}
where it was shown that the Pontryagin index, which is trivially invariant
under small gauge rotations, also does not change under both transformations
(\ref{G11}) and (\ref{G15}), i.e.
\begin{equation}
\nu[A^U] = \nu[(A^U)^V] = \nu[A] \stackrel{(\ref{G9})}{=} n[\Omega]
\label{G13}\,. 
\end{equation}
In the Polyakov gauge, however, the gauge function $\Omega(\vek{x})$ is 
diagonalized and from (\ref{G15}) we infer
\[
n[\Omega] = n[V^{-1}] + n[\omega] + n[V] = n[\omega] = 0 
\label{G13a}
\]
unless the coset transformation $V(\vek{x})$ is singular somewhere on the 
spatial manifold $\mathsf{M}$.
Again we conclude that the winding number of $\Omega$ and thus the 
Pontryagin index in Polyakov gauge arises exclusively from the defects.


\section{Gauge Fixing Defects}
\label{sec:3}

A group element is called irregular when two of its eigenvalues
are degenerate. For irregular $\Omega(\vek{x})$ we can always consider 
the two degenerate eigenvalues of $\Omega(\vek{x})$
to belong to an $SU(2)$ subgroup of the full gauge group $G=SU(N)$.
Therefore it suffices to consider the gauge group $SU(2)$ where the 
irregular elements are given by the centre elements $\Omega = \pm \unit$.
We define an individual defect $\DD_i$, as usual, as a connected set 
of points for which the smooth mapping $\Omega(\vek{x})$ takes on an irregular 
element,
\begin{equation}
\DD_i = \{\vek{x}\in\MM\,\,,\,\,\Omega(\vek{x}) = \mbox{const} = \pm \unit \}
\subseteq \MM\,\,,\qquad \DD_i\,\,\,\mbox{connected}.
\label{G21}
\end{equation}
Since $\Omega(\vek{x})$ is time-independent, all defects are \emph{static}
and it suffices to investigate the three-dimensional space $\MM$.
According to the dimensionality we distinguish the following defects: 
\begin{itemize}
\item $\pi_2(\MM\backslash\DD_i)\neq \emptyset$: Isolated point defects 
     (magnetic monopoles)
\item $\pi_1(\MM\backslash\DD_i)\neq \emptyset$: Closed line defects 
\item $\pi_0(\MM\backslash\DD_i)\neq \emptyset$: Closed domain 
     walls
\end{itemize}
\emph{Open} line and wall defects are topologically equivalent to isolated 
point defects. Similarly, \emph{three-dimensional} defects give merely rise to 
additional internal boundaries of $\MM$ where the gauge function 
$\Omega(\vek{x})$ takes an irregular element. 
The volume of the three-dimensional defects does not contribute to the 
winding number, since $\Omega(\vek{x})$ is a constant $\pm\unit$ there.
Such volume defects can therefore  be treated analogously to the point 
defects and will not be considered here.

\begin{figure}
\begin{centering}
\hfill
\begin{minipage}{5.3cm}
\includegraphics[width = 5cm]{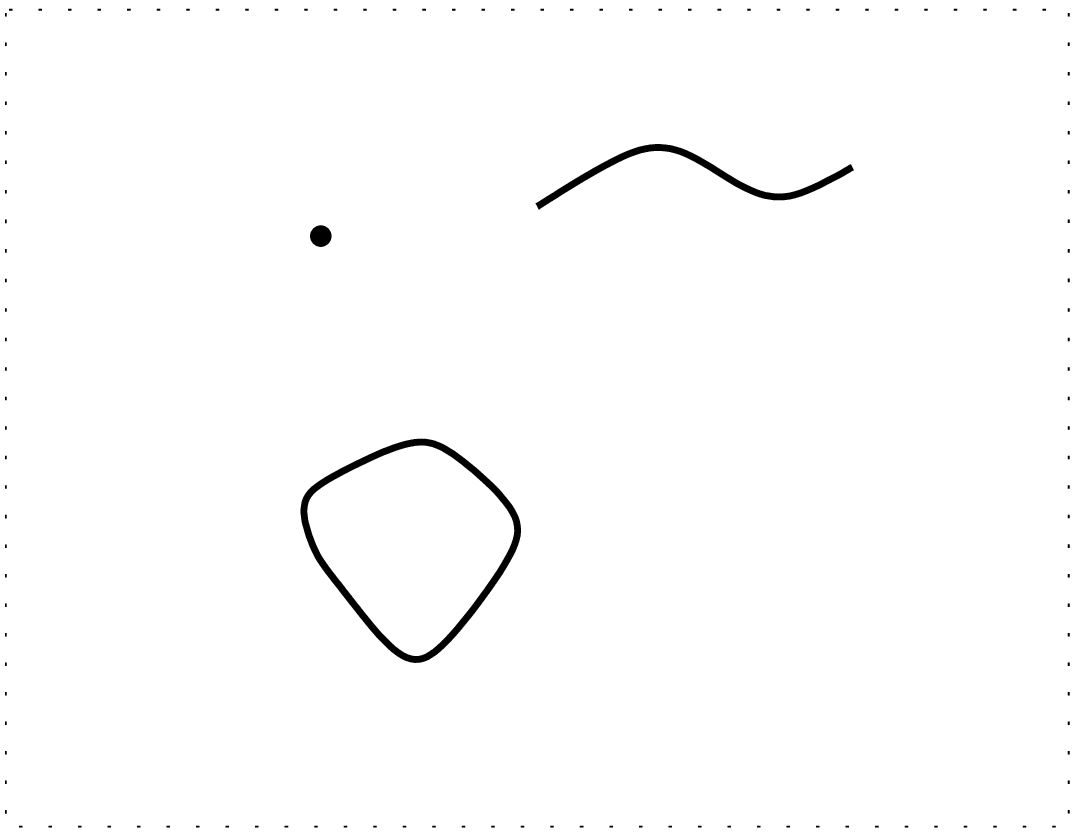}\\
\centerline{(a)}
\end{minipage}
\hfill
\begin{minipage}{5.3cm}
\includegraphics[width = 5cm]{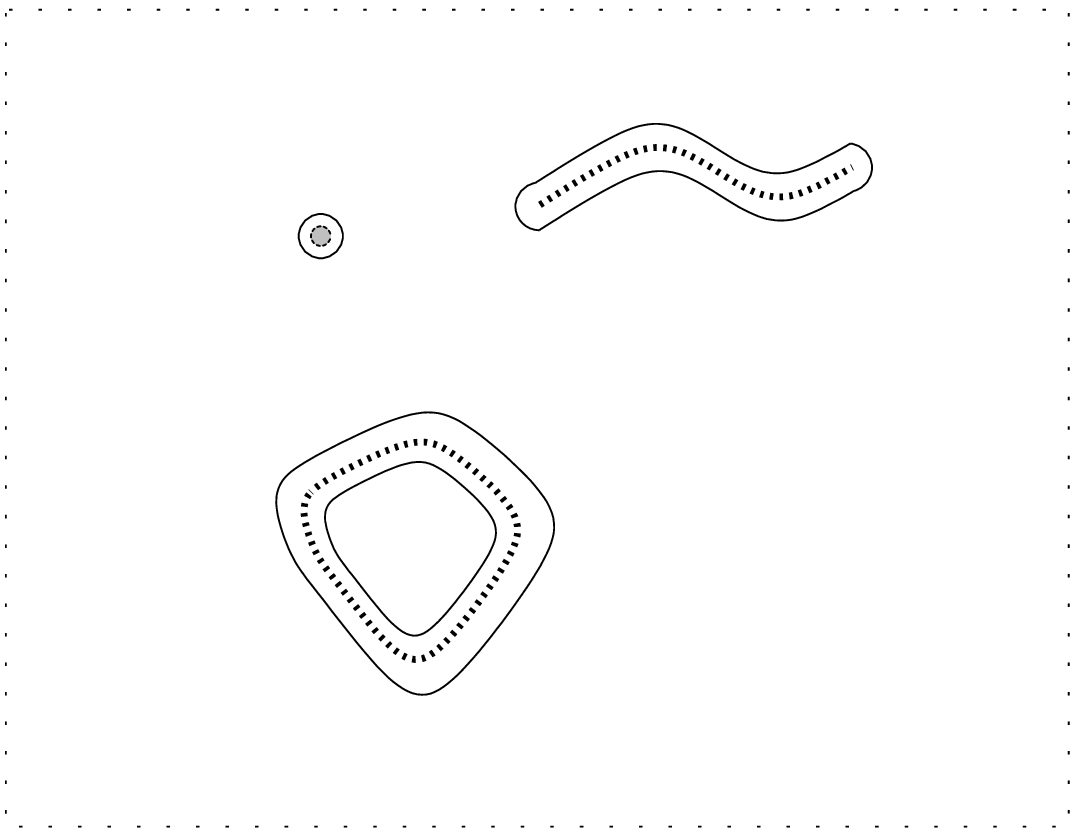}\\
\centerline{(b)}
\end{minipage}
\hfill
\end{centering}
\caption[Wrapping of diagonalisation defects]
{\textsl{Wrapping of some generic diagonalisation defects.
(a) Point defects and open or closed line defects, 
(b) wrapping of these defects by closed surfaces with infinitesimal volume 
   $\epsilon$.}}
\label{fig:1}
\end{figure}

To proceed further we wrap the defects by closed surfaces which are 
infinitesimally close to the defects, see fig.~\ref{fig:1}. The defect $\DD_i$ 
together with its wrapping is denoted by $\DD_i^{\epsilon}$. We then cut out 
the defects together with their wrappings from our spatial manifold $\MM$ giving 
rise to the punctured space
\begin{equation}
\Me = \MM \,\backslash \,\bigcup\limits_i\DD_i^\epsilon
\label{G22}
\end{equation}
where each defect gives rise to an internal surface enclosing the defect, 
see fig.~\ref{fig:1}. We will assume that the punctured space $\Me$ admits 
a covering $\{\XX_\alpha\}$ by closed contractible sets $\XX_\alpha$  
\begin{equation}
\Me = \bigcup\limits_\alpha \XX_\alpha
\label{G23}
\end{equation}
with
\begin{equation}
\XX_\alpha \cap \XX_\beta = \partial\XX_\alpha \cap \partial \XX_\beta
\label{G24}
\end{equation}
such that on each (topologically trivial) patch $\XX_\alpha$
we have smooth diagonalization maps $V_\alpha(\vek{x})$ and 
$\omega_\alpha(\vek{x})$ with 
$\Omega = V_\alpha^{-1}\,\omega_\alpha\,V_\alpha$.
The triangulation (\ref{G23}) implies that, according to (\ref{G24}),
the closed oriented patches $\XX_\alpha$ intersect precisely on their 
boundaries. Note that the patches $\XX_\alpha$ are oriented, 
and so is their intersection. Contrary to ordinary sets, the intersection
operator does therefore not commute, 
\begin{equation}
\XX_\alpha \cap \XX_\beta = - \XX_\beta \cap \XX_\alpha
\label{sonja}\,.
\end{equation}
In the presence of closed domain walls the manifold $\Me$ consists of 
disconnected pieces $\MM_a$, $\MM=\bigcup_a \MM_a$ separated by the domain 
walls. In each connected component $\MM_a$ the induced gauge field 
$a_0=-\frac{1}{\beta}\,\ln\omega(\vek{x})$ can be chosen smoothly, but
some care has to be taken when extending $\omega(\vek{x})$ and 
$\ln\omega(\vek{x})$ over different domains $\MM_a$.

To see this, recall that $\Omega(\vek{x})$ is smooth over the common 
boundary of two patches $\XX_\alpha$ and $\XX_\beta$, and we have 
\begin{equation}
\Omega(\vek{x}) = V^{-1}_\alpha(\vek{x})\,\omega_\alpha(\vek{x})\,V_\alpha(\vek{x})
= V^{-1}_\beta(\vek{x})\,\omega_\beta(\vek{x})\,V_\beta(\vek{x})
\label{G25}
\end{equation}
so that the diagonal maps $\omega_\alpha(\vek{x})$ are related by
\begin{equation}
\omega_\alpha(\vek{x}) = h_{\alpha\beta}(\vek{x})\,\omega_\beta(\vek{x})\,
h_{\alpha\beta}^{-1}(\vek{x})
\label{G26}
\end{equation}
with the \emph{transition functions}
\begin{equation}
h_{\alpha\beta}(\vek{x}) = V_\alpha(\vek{x})\cdot V_\beta^{-1}(\vek{x})
\label{G27}\,.
\end{equation}
They obviously satisfy the co-cycle condition
\begin{equation}
h_{\alpha\beta}\cdot h_{\beta\gamma} = h_{\alpha\gamma}
\label{G28}\,.
\end{equation}
From eq.~(\ref{G26}) we infer that $h_{\alpha\beta}$ takes values in the 
normalizer $N= W \times H$ of the Cartan subgroup $H$ and consequently, 
the diagonalizations $\omega_\alpha$ and $\omega_\beta$ coincide up to 
a Weyl transformation. Since our color group is simply connected,
$\pi_1(SU(N)) = \emptyset$, the picture $P\subset H$ of a Weyl alcove under 
the exponential map represents a fundamental domain for the Cartan group, 
i.e.~any Weyl transformation leads out of $P$. Thus, by restricting our 
diagonalization $\omega$ to $P\subset H$, it is possible to choose 
\begin{equation} 
\omega_\alpha = \omega_\beta
\label{flo}
\end{equation}
smoothly on the overlap of two patches.
Furthermore, since the branch cut of the logarithm is situated at the defects,
which are excluded from our manifold $\Me$, and since the subset $P\subset H$ 
is simply connected, the same is true for the \emph{logarithm} of our 
diagonalization, 
\begin{equation}
\ln\omega_\alpha = \ln\omega_\beta
\label{andy}\,.
\end{equation}  
On the other hand, there are no overlapping patches between different 
domains $\MM_a$ (disconnected by closed domain walls), and eqs.~(\ref{flo}) and (\ref{andy}) do no longer hold
necessarily. In fact, there is an ambiguity in the choice of fundamental
subsets $P\subset H$ for the diagonalization in the various connected 
regions $\MM_a$. As a consequence, the maps $\omega_a$ and $\omega_b$ at 
infinitesimally close points on opposite sides of a closed domain wall are 
related by
\[
\omega_a = \omega_b\qquad\mbox{or}\qquad\omega_a = \omega_b^\dagger
\label{daniel}
\]
and
\[
\ln\omega_a = \pm\ln\omega_b + 2 \pi i k \sigma_3\,,\qquad k\in\Integer
\label{elisabeth}\,.
\]


\section{The Winding Number Expressed by Defects}
\label{sec:4}
The integrand in the winding number $n[\Omega]$  can be locally (i.e.~within 
a patch $\XX_\alpha$) expressed as a total derivative 
\begin{equation}
\tr\,(L\wedge L\wedge L) = d\,\GG(\omega,V_\alpha)
\label{G50}
\end{equation}
where
\begin{equation}
\mathsf{G}[\omega,V_\alpha] = - 6\,\tr\left(\mathcal{A}_\alpha \wedge 
\omega\,d\omega^{-1}\right) + 3\,\tr \left(\mathcal{A}_\alpha\,\omega^{-1}
\wedge \mathcal{A}_\alpha \omega \right)\,\,,\qquad
\mathcal{A}_\alpha = V_\alpha\,d V_\alpha^{-1}
\label{G51}.
\end{equation}
Applying Stokes' theorem, we find
\begin{equation}
n[\Omega] = - \frac{1}{24 \pi^2} \sum\limits_\alpha 
\int\limits_{\partial \XX_\alpha} \GG[\omega,V_\alpha]
\label{G52}\,.
\end{equation}
A patch $\XX_\alpha$ can have a common border with another patch $\XX_\beta$ or 
with a defect $\DD_i^\epsilon$ whence we find for the surface of a 
patch\footnote{Note that this equation defines an orientation for the 
intersection operator of the (oriented) patches.}
\begin{equation}
\partial\XX_\alpha = \sum\limits_{\beta\neq \alpha} \XX_\alpha \cap \XX_\beta + 
\sum\limits_i \XX_\alpha\cap \DD_i^\epsilon
\label{G38}\,.
\end{equation}
Using this decomposition of the surface $\partial\XX_\alpha$ we
obtain 
\begin{equation}
n[\Omega] = - \frac{1}{48 \pi^2} \sum\limits_{\alpha,\beta}\,
\int\limits_{\XX_\alpha\cap\XX_\beta}\left(\GG[\omega,V_\alpha] - 
\GG[\omega,V_\beta]\right) - \frac{1}{24 \pi^2}\sum\limits_{\alpha,i}\,
\int\limits_{\XX_\alpha\cap\DD_i^\epsilon} \GG[\omega,V_\alpha]
\label{G53}\,,
\end{equation}
where we exploited the different orientation (\ref{sonja}) of the intersection 
$\XX_\alpha\cap\XX_\beta$, as seen from $\XX_\alpha$ and $\XX_\beta$,
respectively. From (\ref{G51}) and (\ref{G27}) we find 
\begin{equation}
\GG[\omega,V_\alpha]-\GG[\omega,V_\beta]=-6\, d\,\tr\left(h_{\alpha\beta}d
h_{\alpha\beta}^\dagger \ln\omega\right)
\label{alex1} .
\end{equation}
Using the cocycle condition (\ref{G28}), the first integral in
(\ref{G53}) can be rewritten by means of Stokes' theorem,
\begin{equation}
\int\limits_{\XX_\alpha\cap\XX_\beta}\left(\GG[\omega,V_\alpha] - 
\GG[\omega,V_\beta]\right)=-6\sum_i 
\int\limits_{\XX_\alpha\cap\XX_\beta\cap\DD_i^\epsilon}\tr\left(h_{\alpha\beta}
d h_{\alpha\beta}^\dagger \ln\omega\right) 
\label{alex2} .
\end{equation}
Here we have decomposed the boundary 
of $\XX_\alpha \cap \XX_\beta$ as (cf.~eq.~(\ref{G38}))
\begin{equation}
\partial(\XX_\alpha\cap\XX_\beta) = \bigcup\limits_\gamma \XX_\alpha\cap\XX_\beta
\cap\XX_\gamma + \bigcup\limits_i \XX_\alpha\cap\XX_\beta\cap\DD_i^\epsilon
\label{G45}\,.
\end{equation}
Turning to the second integral in (\ref{G53}) we observe that the last term 
in eq.~(\ref{G51}) does not contribute, since it
vanishes for $\epsilon\to 0$, i.e.\ $\omega\to\pm\unit$. On the other hand the
first term in (\ref{G51}) can be written as
\begin{equation}
- 6\,\tr\left(\mathcal{A}_\alpha \wedge\omega\,d\omega^{-1}\right)
=-6\,d\, \tr\left(\mathcal{A}_\alpha \ln\omega\right)+6\,\tr\left(d\mathcal{A}_\alpha 
\ln\omega\right)
\label{alex3} .
\end{equation}
Hence by using Stokes' theorem, we obtain for the second integral in 
(\ref{G53})
\begin{equation}
\sum\limits_{\alpha,i} \int\limits_{\XX_\alpha\cap\DD_i^\epsilon} 
\GG[\omega,V_\alpha] = - 6 \sum\limits_{\alpha,i} 
\int\limits_{\partial(\XX_\alpha\cap\DD_i^\epsilon)} \tr\left(\mathcal{A}_\alpha
\,\ln\omega\right) + 6 \sum\limits_{\alpha,i} 
\int\limits_{\XX_\alpha\cap\DD_i^\epsilon} \tr\left(d \mathcal{A}_\alpha\,\ln
\omega\right)
\label{katja}\,.
\end{equation}
Expressing the surface $\partial(\XX_\alpha\cap\DD_i^\epsilon)$ analoguously
to eq.~(\ref{G45}) and taking care of the proper orientation of the 
intersections (cf.~(\ref{sonja})), we have 
\begin{equation}
\partial(\XX_\alpha\cap\DD_i^\epsilon) = - \sum\limits_{\beta} \XX_\alpha\cap
\XX_\beta\cap\DD_i^\epsilon
\label{natalie}\,,
\end{equation}
and the first term on the r.h.s.~of (\ref{katja}) becomes\footnote{The relative
sign in the last integral is again due to the opposite orientation of the
common boundary, as seen from the adjacent patches.} 
\begin{equation}
- 6 \sum\limits_{\alpha,i}\int\limits_{\partial(\XX_\alpha\cap\DD_i^\epsilon)} 
\tr\left(\mathcal{A}_\alpha\,\ln\omega\right) = 
+ 3 \sum\limits_{\alpha,\beta,i}\int\limits_{\XX_\alpha\cap\XX_\beta\cap
\DD_i^\epsilon}\tr\left((\mathcal{A}_\alpha-\mathcal{A}_\beta)
\,\ln\omega\right)
\label{vanessa} \,.
\end{equation}
Using eq.~(\ref{G27}) this term is seen to cancel eq.~(\ref{alex2}) so that
the winding number (\ref{G53}) receives a non-vanishing contribution only from 
the second term in (\ref{katja}),
\begin{equation}
n[\Omega] = - \frac{1}{4\pi^2} \sum\limits_{\alpha,i}\,
\int\limits_{\XX_\alpha\cap\DD_i^\epsilon}\tr\left(\ln\omega\cdot 
T^3\right)\cdot d \mathcal{A}^3_\alpha\,\,,\qquad 
\mathcal{A}_\alpha = V_\alpha\cdot d V_\alpha^{-1}
\label{G54}.
\end{equation}
On a defect $\DD_i$, $\Omega(\vek{x})$ and hence $\omega(\vek{x})$ are 
irregular i.e.~$\omega(\vek{x})$ takes values $(-\unit)^{n_i}=\pm\unit$ 
(for $G=SU(2)$) and is constant. Hence we obtain 
\begin{equation}
\left.\tr\left(\ln\omega\cdot T^3\right)\right|_{\DD_i^\epsilon} =  \pi\,\len_i
\label{G55}\,.
\end{equation}
Furthermore the quantity
\begin{equation}
m_i = \frac{1}{4 \pi}\sum\limits_{\alpha}\int\limits_{\XX_\alpha\cap 
\DD_i^\epsilon} d \mathcal{A}_\alpha^3
\label{G56}
\end{equation}
is the magnetic flux through the wrapping surface of the defect\footnote{Note 
that the normal vector on the wrapping surface around the defect 
$\DD_i^\epsilon$ points out of the punctured space $\Me$, i.e.~towards the 
defect. By contrast, the intersection $\XX_\alpha\cap \DD_i^\epsilon$ is
oriented opposite to $\partial \DD_i^\epsilon$ (cf.~eq.~(\ref{G38})), so that 
our definition (\ref{G56}) yields the usual sign of the magnetic charge, 
i.e.~the magnetic flux \emph{emanating} from the defect.}
and hence represents the magnetic charge of the defect. Applying again 
Stokes' theorem and eq.~(\ref{natalie}) the flux can be expressed as
\begin{equation}
m_i = -\frac{1}{8\pi}\sum\limits_{\alpha,\beta}\,\int\limits_{\XX_\alpha\cap
\XX_\beta\cap\DD_i^\epsilon}\left(\mathcal{A}^3[V_\alpha] - 
\mathcal{A}^3[V_\beta]\right) = \frac{1}{4 \pi} \sum\limits_{\alpha,\beta}\,
\int\limits_{\XX_\alpha\cap\XX_\beta\cap\DD_i^\epsilon}
\tr\left(h_{\alpha\beta}\,d h_{\alpha\beta}^{-1}\cdot T^3\right)
\label{G57}.
\end{equation}
Inserting  eq.~(\ref{G55}) and (\ref{G56}) into equation (\ref{G54})  we obtain 
for the winding number
\begin{equation}
n[\Omega] = - \sum\limits_{i} \len_i\, m_i\,.
\label{G58}
\end{equation}
This is precisely the result derived in ref.~\cite{rei} for a compact spatial
manifold $\MM$. Let us stress that, as the above derivation reveals, all 
defects with $n_i \neq 0$ carrying non-zero magnetic charge $m_i \neq 0$ 
contribute to the topological charge.

Since a shift of $\ln\omega$ by $2\pi i$ leads to the same $\omega(\vek{x})$,
we have a freedom in the choice of $\chi= - i \ln \omega$.
Choosing $\chi$ smooth at the defect where $\omega=\pm\unit$ will in general 
lead $\chi$ outside the first Weyl alcove. As we will illustrate below in this 
case closed domain walls will carry no magnetic charge and hence do not 
contribute to the winding number.

Alternatively we can restrict $\chi$ to the first Weyl alcove 
$\chi\in[0,\pi]$. Then at the defect the integer defined in equation
(\ref{G55}) is restricted to $\len_i=0,1$ and equation (\ref{G58}) becomes
\begin{equation}
n[\Omega] = - \sum\limits_{i\atop{\Omega = -\unit}} m_i
\label{G59}\,\,,
\end{equation}
i.e.~only the defects with $\Omega=-\unit$ contribute. 
Restricting $\chi$ to the first Weyl alcove $\chi\in[0,\pi]$ implies that 
$\chi$ is continuous but not necessarily smooth at the defect, see 
fig.~\ref{fig:3}. In this case closed domain walls with $\omega=-\unit$ 
now carry twice the magnetic charge of a monopole and hence 
contribute to $n[\Omega]$ as will be illustrated below.

Thus in the generic case where only magnetic monopoles and domain walls are
present, eq.~(\ref{G59}) can be more explicitly written as
\begin{equation}
n[\Omega]=- \hspace{-5mm}\sum_{{k \mathrm{(magnetic\, monopoles)}\atop 
\Omega=-\unit}}\hspace{-9mm}m_k \quad-
\sum_{{k \mathrm{(domain\, walls)}\atop \Omega=-\unit}}\hspace{-5mm} m_k
\label{B1}\,.
\end{equation}

We observe that for the smooth parameterization the magnetic monopoles fully 
account for the topological charge (\ref{G58}), while there is an extra 
contribution in eq.~(\ref{B1}) due to magnetically charged domain walls.
In refs.~\cite{wipf,lenz} this domain wall contribution to $n[\Omega]$ was not 
included although $\chi=-i\ln\omega$ was restricted to first Weyl alcove.


\section{Hedgehog Field as Generic Example} 

\begin{figure}
\begin{centering}
\hfill
\includegraphics[width = 4.5cm]{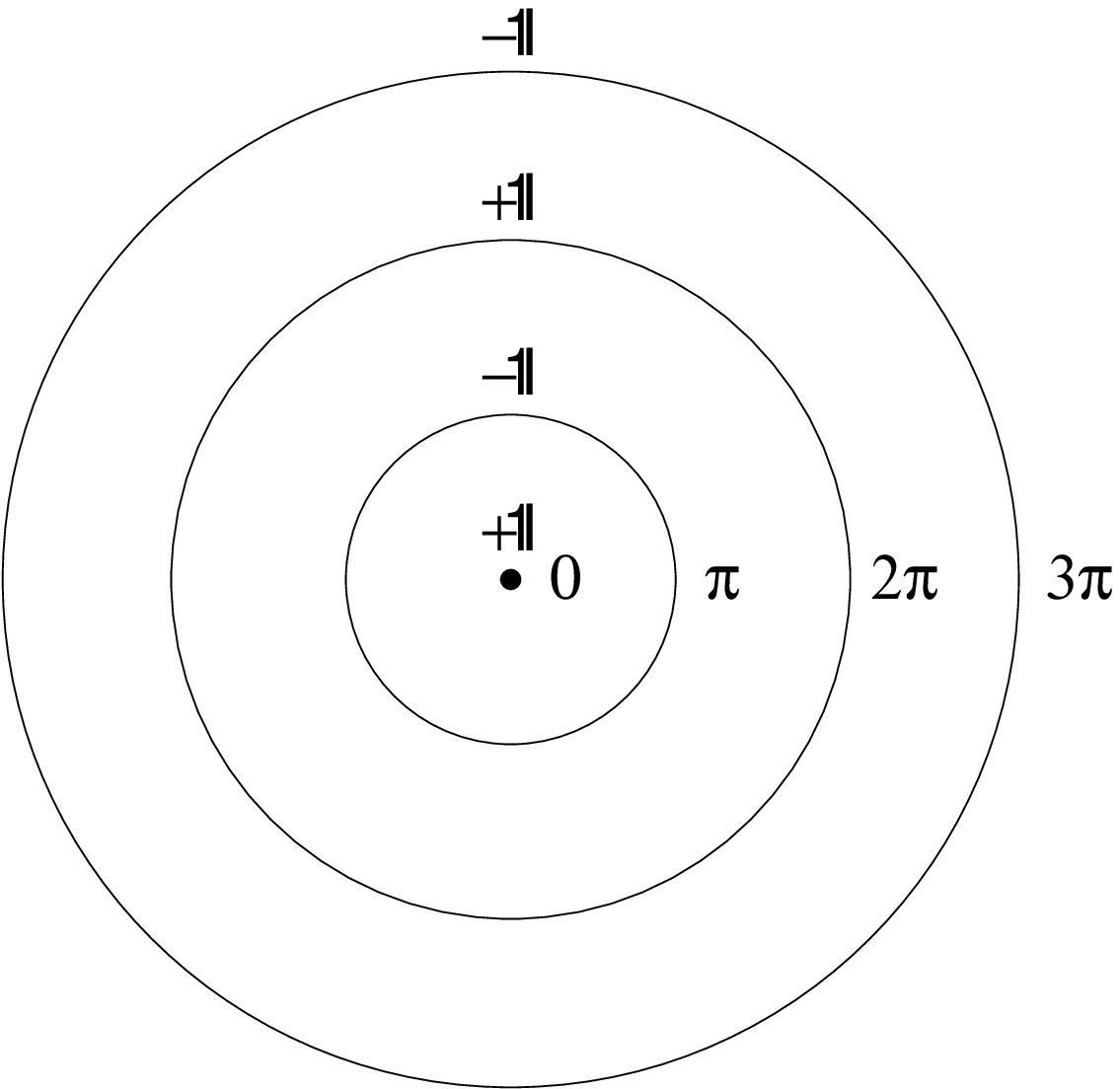} 
\hfill
\includegraphics[width = 5.0cm]{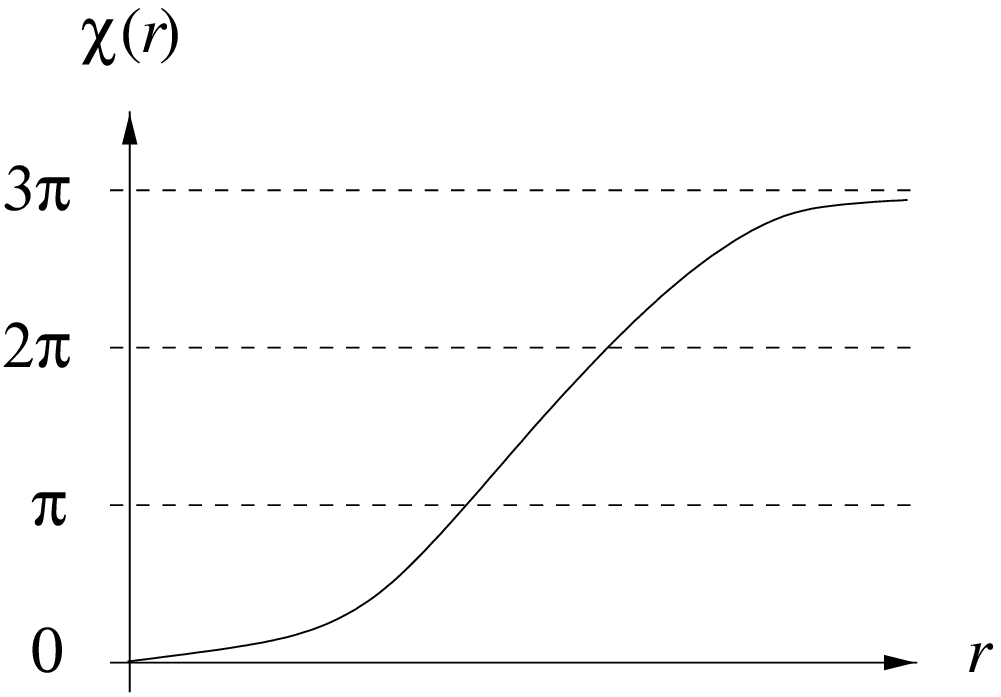}
\hfill
\end{centering}
\caption[Smooth assignment of Weyl alcoves.]
{\textsl{Smooth assignment of alcoves for a hedgehog type of 
mapping $\Omega\,:\,\dot{\Real}^3 \mapsto SU(2)$.
(a) Defect structure of $\Omega$. 
(b) Profile $\chi(r)$ ($\ln\omega = -i \chi \sigma_3$)
    as a function of $r = |\vek{x}|$. 
    The Weyl alcove is changed between the defects, such that $\chi(r)$ becomes
    globally smooth.}}  
\label{fig:2}
\end{figure}

\begin{figure}
\begin{centering}
\hfill
\includegraphics[width = 4.5cm]{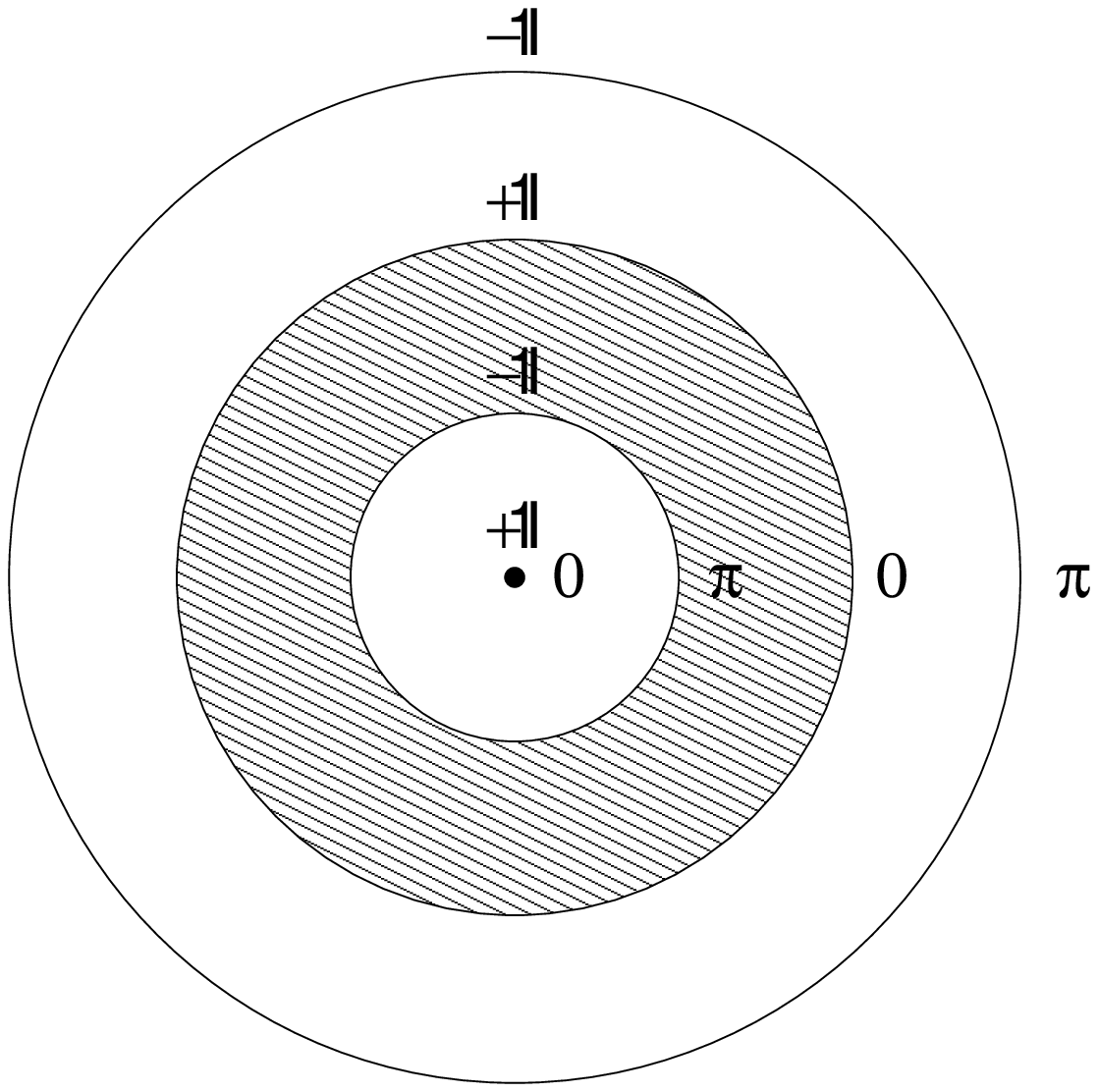}
\hfill
\includegraphics[width = 5.0cm]{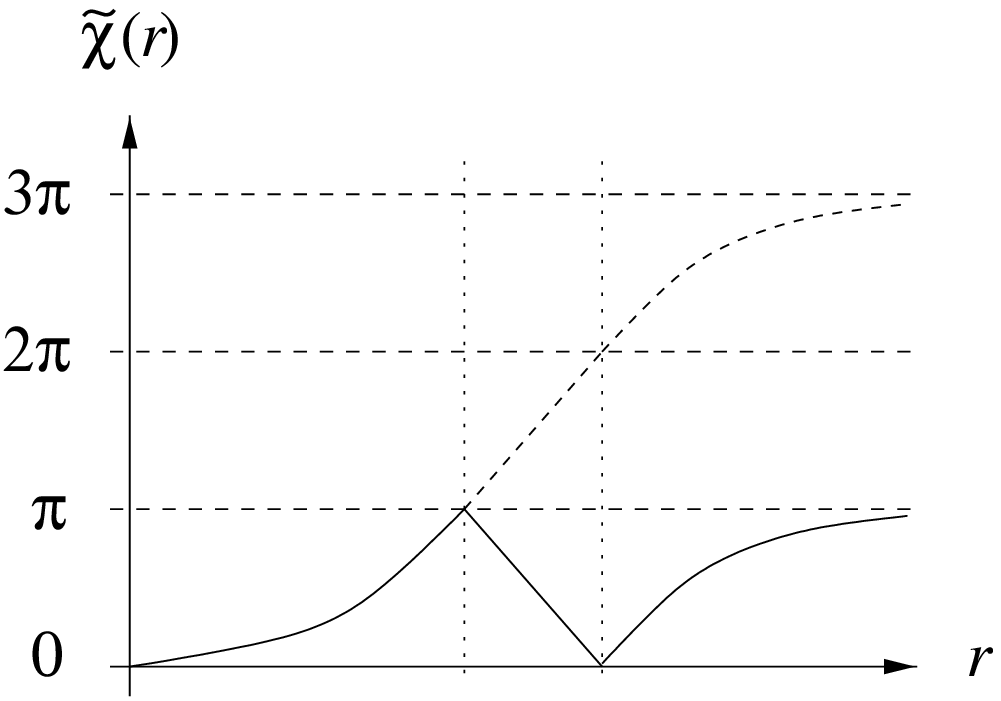}
\hfill
\end{centering}
\caption[Fixed assignment of Weyl alcoves.]
{\textsl{Fixed assignment of alcoves for a hedgehog type of 
mapping $\Omega\,:\,\dot{\Real}^3 \mapsto SU(2)$.
(a) Defect structure of $\Omega$. 
(b) Profile $\tilde{\chi}(r)$ ($\ln\omega = -i \tilde{\chi} \sigma_3$)
    as a function of $r = |\vek{x}|$. 
    The Weyl alcove is rigidly fixed to $\tilde{\chi}\in[0,\pi]$ such that 
    $\tilde{\chi}(r)$ is reflected at the alcove boundary, i.e.~the profile is 
    continuous, but not smooth.}}
\label{fig:3}
\end{figure}

Let us finally illustrate our result for the two different methods 
eqs.~(\ref{G58}) and (\ref{G59})
by means of a specific example. The prototype of smooth maps 
$\Omega\,:\,\dot{\Real}^3\to SU(2)$ with non-vanishing winding number
is provided by the well-known \emph{hedgehog} configuration,
\begin{equation}
\Omega(\vek{x}) = \exp\left(i \chi(r) \cdot\hat{\vek{x}}\Vek{\sigma}\right)
= \unit\cdot\cos\chi(r) +  i\,\hat{\vek{x}}\Vek{\sigma}\cdot\sin\chi(r)
\,\, ;\qquad r \equiv |\vek{x}| 
\label{G60}\,.
\end{equation}  
For this map to be smooth, we have to avoid the singularity at the origin
where $\hat{\vek{x}}$ is ill-defined, 
\begin{equation}
\sin\chi(0) = 0 \qquad \iff \qquad \chi(0) = n_0\cdot\pi \,\, ; \qquad
n_0\in\Integer
\label{G61}\,.
\end{equation}
Furthermore, the compactification $\Real^3\to\dot{\Real}^3$ implies that
$\Omega(\vek{x})$ be angle-independent as $r = |\vek{x}| \to \infty$, i.e.~the
$\hat{\vek{x}}$-dependent part of $\Omega$ must vanish as $r\to\infty$,
whence
\begin{equation}
\sin\chi(\infty) = 0 \qquad \iff \qquad \chi(\infty) = n_\infty\cdot\pi\,\,;
\qquad n_\infty\in\Integer 
\label{G62}\,.
\end{equation}
The winding number of such a hedgehog map is determined by the
difference of the two integers from the profile boundary conditions via 
$n[\Omega]=n_\infty -n_0$. 
For definiteness we choose 
\begin{equation}
n_0=0\,, \quad\mbox{i.e.}\quad n_\infty=n[\Omega]
\label{B2}\,.
\end{equation}
As expected, there is a 
continuous diagonalisation of this map, 
\begin{equation}
\omega(\vek{x}) = \omega(r) = \exp\left(i\tilde{\chi}(r)\,\sigma_3\right)
\label{G64}
\end{equation} 
where the profile $\tilde{\chi}$ reflects our choice of Weyl alcoves in 
the connected regions separated by the domain walls. There are two basically
different choices: 
\begin{enumerate}
\item Choose the alcoves in every connected region such that $\tilde{\chi}(r)$
      is globally smooth, i.e.~$\tilde{\chi}(r) = \chi(r)$, see 
      fig.~\ref{fig:2}.  
\item Exploit the arbitrariness $\chi\to\chi+2\pi$ and the Weyl symmetry 
     $\chi\to - \chi$ to fix the Weyl alcove $\tilde{\chi}\in[0,\pi]$ for all 
     connected regions of $\dot{\Real}^3_\epsilon$, i.e.~$\tilde{\chi}(r)$ is 
     the profile obtained from the original $\chi(r)$ by \emph{reflections} at 
     the boundary of the alcove, see fig.~\ref{fig:3}. While both
     $\omega(\vek{x})$ and $\ln\omega(\vek{x})$ are still continuous, 
     they fail to be smooth across the domain walls. 
\end{enumerate}
Note that we may enclose both the monopoles and the domain walls by simple
$2$-spheres $S^2$ so that it is convenient to switch to spherical coordinates 
$\vek{x} \to (r,\vartheta,\varphi)$ on $\dot{\Real}^3$. We may cover
every connected region in the punctured space $\dot{\Real}^3_\epsilon$ 
by just two coordinate patches $\XX_{\pm}$, so that their intersections with the 
wrapping sphere $S^2$ yield the northern ($S^2_{+}$) and southern 
($S^2_{-}$) hemisphere. In each of these contractible patches, we find a 
smooth diagonalising coset lift $V_{\pm}(r,\vartheta,\varphi)$. 

\begin{figure}
\begin{centering}
\hfill
\begin{minipage}[t]{6.5cm}
\includegraphics[width=6cm]{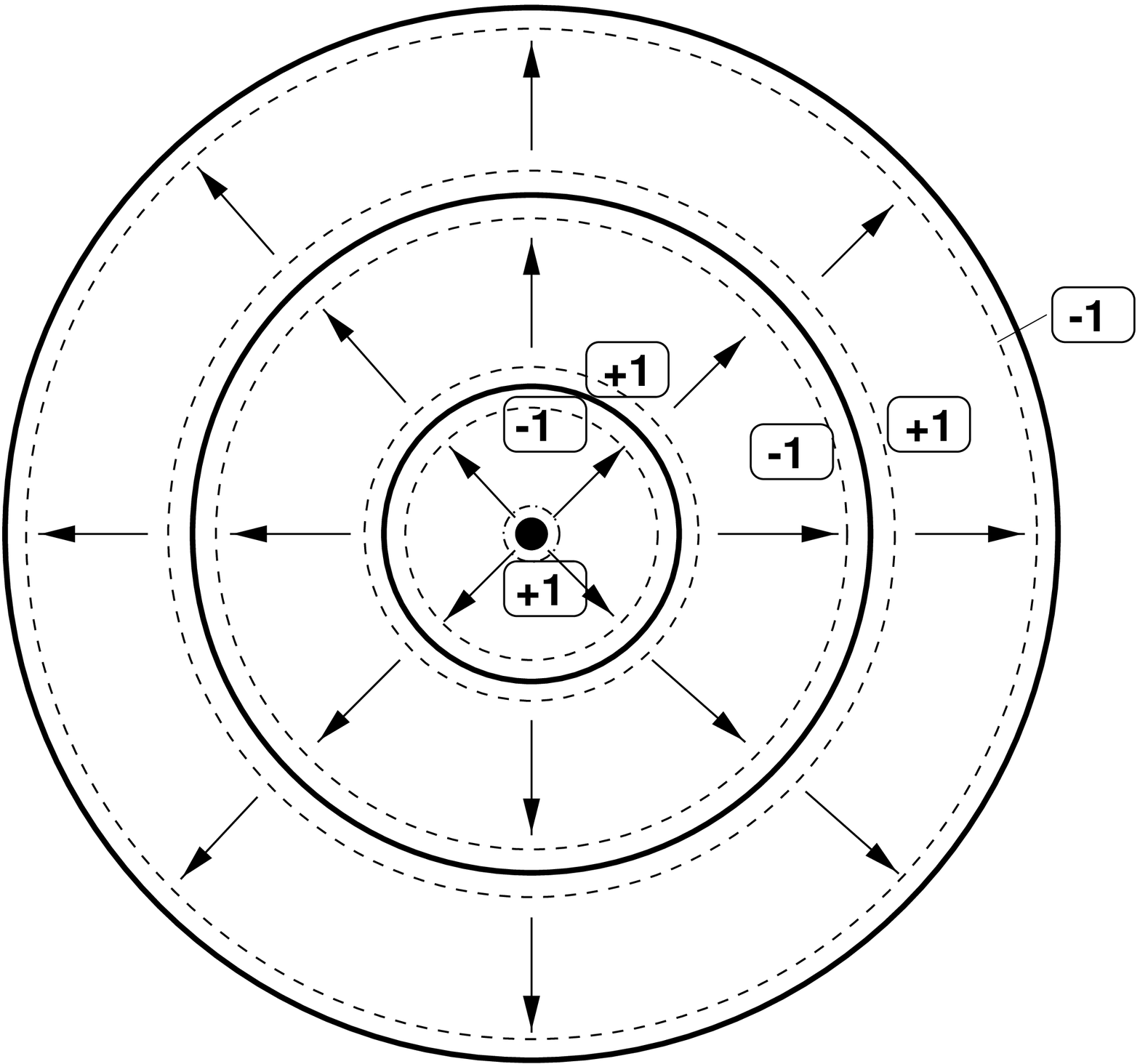}
\end{minipage}
\hfill
\begin{minipage}[t]{6.5cm}
\includegraphics[width=6cm]{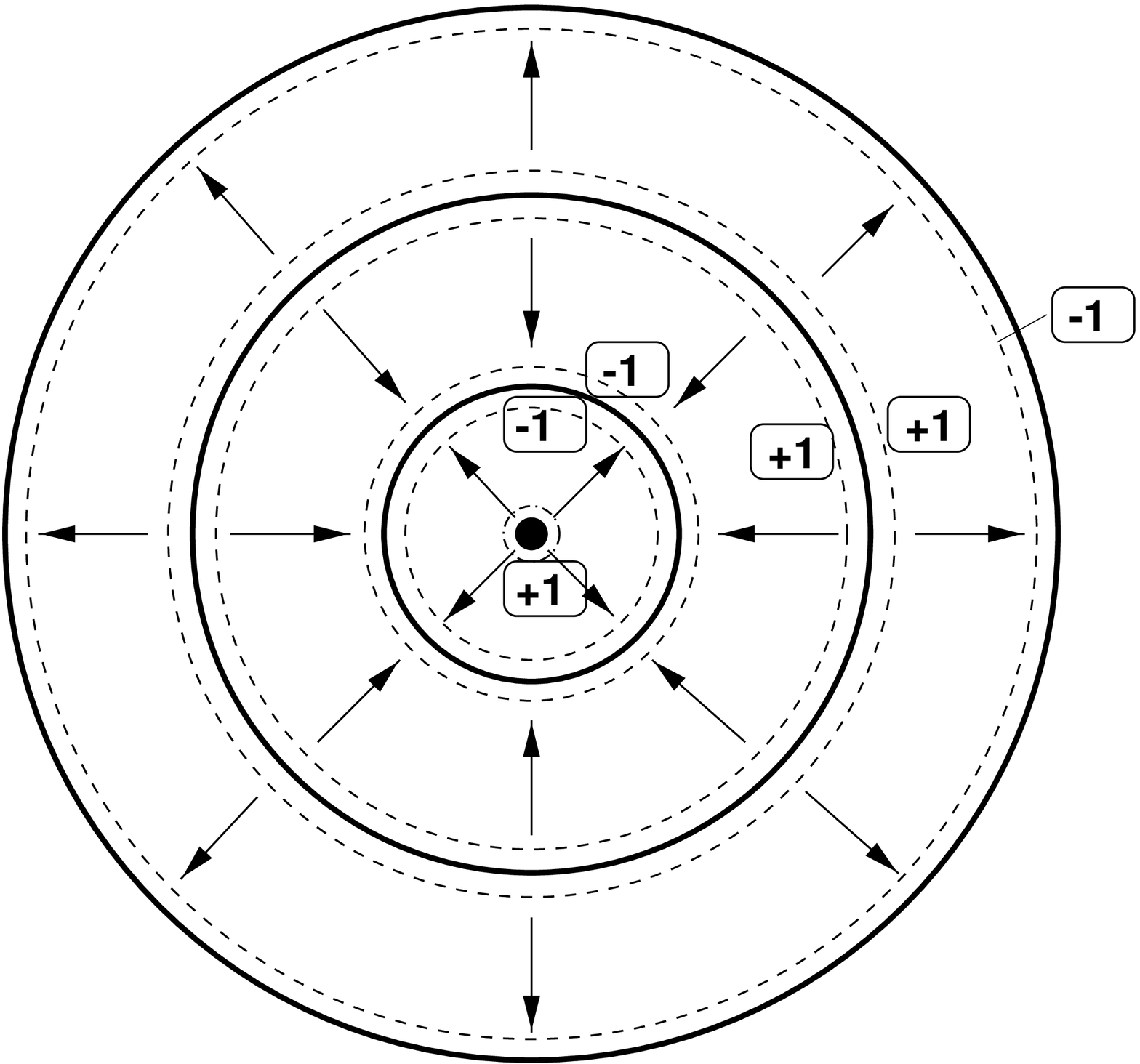}
\end{minipage}
\hfill
\caption{\label{fig:kandinsky}
\textsl{The $n=3$ hedgehog mapping with continuous profile (left) 
and restricted profile (right). 
Solid circles represent domain wall defects, dashed circles represent
the wrapping surfaces and the dot in the origin symbolises the magnetic 
monopole. The numbers in the boxes denote the magnetic charge of the 
respective wrapping, which is determined from the relative orientation of the
surface (not indicated for clearity) and the magnetic field (the radial 
arrows). }} 
\end{centering}
\end{figure}

In the case of a smooth choice of $\chi(r)$ (see fig.~\ref{fig:2}),
the hedgehog $\Omega$ can be diagonalized
by the \emph{same} set of coset lifts $V_\pm$ for all $r$. An explicit form is 
given by \cite{rei}, 
\begin{equation}
V_{+}(\vek{x}) = \exp(i\frac{\vartheta}{2}\,\vek{e}_\varphi\Vek{\sigma})
\,\, ; \qquad
V_{-}(\vek{x}) = h_{\pm}(\vek{x}) \cdot V_{+}(\vek{x})\,,
\label{G65}\nonumber 
\end{equation}
where the transition function 
\[h_{\pm} = \exp\left(i\varphi\,\sigma_3\right)\]
has support in the overlap between 
the two hemispheres, i.e.~on the equator $S^1$ defined by 
$\vartheta = \frac{\pi}{2}$. From these expressions, we can easily calculate 
the magnetic flux of the induced Abelian potential 
$\mathcal{A}^3_{\pm} = \left(V_{\pm}\,d V^{-1}_{\pm}\right)^3$. 
The magnetic field is always directed radially outwards (pointing to 
infinity), and to every wrapping surface, we assign a 
magnetic charge (\ref{G56})
\[
m = -\frac{1}{4\pi}\sum\limits_{\pm}\int_{S^2_{\pm}}d \mathcal{A}^3_{\pm}
= -\frac{1}{4\pi}\int\limits_{S^2_{+}\cap S^2_{-}} \left(\mathcal{A}^3_{+} - 
\mathcal{A}^3_{-}\right) = \frac{1}{2\pi}\int\limits_{{\rm equator\,} S^1} 
d\varphi= \pm 1
\label{G66}\,.  
\]
Recall that the sign of this charge is determined by the orientation of the 
intersection $\XX_\alpha\cap \DD_i^\epsilon$, which is \emph{opposite} to the 
orientation of the wrapping surface. 
Thus, if the surface $\partial\DD_i^\epsilon$ encloses the defect from the 
outside (i.e.~it is closer to infinity than the defect), the intersection
$\XX_\alpha\cap\DD_i^\epsilon$ is aligned with the magnetic field whence the 
defect has charge $(+1)$. On the other hand, a wrapping surface 
closer to the origin than the defect is given a charge $(-1)$ 
(see also fig.~\ref{fig:kandinsky}).

Thus, for a smooth choice of profile $\chi(r)$ (see figs.~\ref{fig:2} and
\ref{fig:kandinsky} (left)), the magnetic charges of the two surfaces $S^2$ 
wrapping the domain wall cancel and there is no net (intrinsic) magnetic charge 
on the domain walls. Hence the magnetic field goes smoothly through the domain 
wall without noticing its existence. Thus for a smooth choice of $\chi(r)$ the 
domain walls do not contribute to the Pontryagin index. The Pontryagin index is 
entirely determined by the two monopoles at $r=0$ and $r=\infty$.
In fact since $\chi(0) = 0$, only 
the monopole at infinity contributes. From $\chi(\infty) = n[\Omega]\,\pi$
and the magnetic charge $m_\infty = -1$ of this monopole, eq.~(\ref{G58})
yields the correct winding number.

Consider now the alternative case of a fixed assignment of alcoves as in 
fig.~\ref{fig:3}. In order to reflect $\tilde{\chi}(r)$
in the alcove $[0,\pi]$, we need to combine the coset lift $V_\pm$ from 
the smooth case above with a Weyl flip in the transition functions 
$h_\pm$. This in turn leads to a change in the orientation of 
the magnetic field inside the shaded region of fig.~\ref{fig:3}:
In this domain, the field is directed radially inwards, i.e.~towards the 
origin, while we have the usual (outward) orientation in the remaining space. 
The orientation of the wrapping surfaces is still determined by a normal 
vector pointing towards the defect. Taking the specific example $n=3$ for 
simplicity (see figure \ref{fig:kandinsky} (right)), we encounter the following 
defects:
\begin{itemize}
\item The monopole in the origin is a $(+\unit)$-defect and does not 
      contribute to $n[\Omega]$.
\item The first domain wall has two wrapping surfaces. The inner one
      is directed outwards (towards the domain wall) and this coincides with 
      the orientation of the magnetic field. From the rules explained above, 
      it is assigned a charge $m^{-} = -1$.
      The outer wrapping surface is directed inwards (towards the defect) and 
      this \emph{also} coincides with the flipped orientation of the magnetic
      field in the shaded region. It thus also carries a charge 
      $m^{+} = -1$.
      Altogether, the first domain wall hence carries a total  magnetic 
      charge $m = -2$. 
\item The next domain wall is a $(+\unit)$-defect and again does not 
      contribute. However, it carries magnetic charge $+2$ and flips the 
      magnetic field back to pointing outwards.
\item The monopole at infinity is a $\Omega = - \unit$ defect and has charge 
      $(-1)$ since the magnetic field has again its standard orientation 
      pointing to infinity.     
\end{itemize}  
With these observations, our formula (\ref{B1}) gives the correct result
\[
n[\Omega] = -\left[(-2) \,+\,(-1)\right] = +3
\label{G69}\,. 
\] 
It should be noted that the similar analysis of refs.~\cite{wipf,lenz}, did not 
take domains walls into account, and thus erroneously predicts $n[\Omega] = 1$
for the presently considered example.


\section{Summary and Conclusions}
We have investigated the topological charge of Yang-Mills fields in 
Polyakov gauge. Our main results are given by eqs.~(\ref{G58}) and 
(\ref{G59}) and can be summarized as follows: If the temporal gauge field 
$A_0(x)$ is chosen smoothly as in \cite{rei}, the Pontryagin index of a generic
field configuration is entirely 
given by magnetic monopoles. On the other hand, if 
$\chi = \beta A_0$ is restricted to the first Weyl alcove, in addition 
magnetic charges for the closed domain walls arise which also contribute to the 
Pontryagin index, see eq.~(\ref{B1}). 

Other, open, magnetically charged defects, like open domain walls or lines, 
are topologically equivalent to magnetic monopoles and can be treated in the 
same way. 

Although these results have been obtained in Polyakov gauge, we believe 
that they are generic for all Abelian gauges. 
In fact, recent lattice calculations performed in the maximum Abelian gauge
\cite{markum} show also clear correlations between the (topological charge of the)
instantons and magnetic monopoles.\\[3mm]
\textbf{Acknowledgement:}\\[0.5mm]
We thank M.~Engelhardt for a careful reading of the manuscript and critical
remarks. 




\begin{thebibliography}{99}
\bibitem{bali}
       {G.S.~Bali, V.~Bornyakov, M.~Mueller-Preussker, K.~Schilling,
       Phys.~Rev.~\textbf{D54} (1996) 2863.}      
\bibitem{thooft}
       {G.~{}'t Hooft, Nucl.~Phys.~\textbf{B190}[FS3] (1981) 455.}       
\bibitem{giacomo}
       {A.~Di Giacomo, \textsl{talk presented at QCD '98 (Montpellier)}
        hep-lat/9809014.}
\bibitem{cash}
       {T.~Banks, A.~Casher, Nucl.~Phys.~\textbf{B169} (1980) 103.}
\bibitem{rei}               
       {H.~Reinhardt, Nucl.~Phys.~\textbf{B503} (1997) 505.}
\bibitem{wipf}
       {C.~Ford, U.~G.~Mitreuter, T.~Tok, A.~Wipf, J.M.~Pawlowski,
        \textsl{Monopoles, Polyakov-Loops and Gauge Fixing on the 
        Torus}, hep-th/9802191.}
\bibitem{lenz}
       {O.~Jahn, F.~Lenz, \textsl{Structure and Dynamics of Monopoles 
        in Axial-Gauge QCD}, hep-th/9803177.}
\bibitem{jackiw}
       {R.~Jackiw, Rev.~Mod.~Phys.~\textbf{52/4} (1980) 661.}
\bibitem{johns}
       {K.~Johnson, L.~Lellouch, J.~Polonyi, Nucl.~Phys.~\textbf{B367}
        (1991) 675.}
\bibitem{hugo}
       {H.~Reinhardt, Mod.~Phys.~Lett.~\textbf{A11} (1996) 2451.}
\bibitem{blau}               
        {M.~Blau, G.~Thompson, Comm.~Math.~Phys.~\textbf{171} (1995) 639.}
\bibitem{markum}
        {H.~Markum, W.~Sakuler, S.~Thurner, 
         Nucl.~Phys.~Proc.~Suppl.~\textbf{47} (1996) 254.\\
         M.~Feuerstein, H.~Markum, S.~Thurner, \textsl{talk presented at
         Hirschegg '97}, hep-lat/9702006.}
\end{thebibliography}
\end{document}